\documentclass[fleqn,twocolumn]{openjournal}

\usepackage{mathptmx}
\usepackage[T1]{fontenc}
\usepackage{ae,aecompl}
\usepackage{graphicx}	
\usepackage{amsmath}	
\usepackage{amssymb}	
\usepackage{latexsym}
\usepackage{longtable} 
\usepackage[figuresleft]{rotating}
\usepackage{epsf}
\usepackage{hyperref}
\hypersetup{colorlinks=true,
	urlcolor=blue,  
	linkcolor=red,  
	citecolor=blue, 
    menucolor=blue, 
    urlcolor=blue}  

\begin{document}

\title{\large{\textbf{The Millions of Optical - Radio/X-ray Associations (MORX) Catalogue, \symbol{118}2}}\vspace{15pt}} 
\author{Eric Wim Flesch}
\affiliation{PO Box 15, Dannevirke 4942, New Zealand; eric@flesch.org} 
\email{eric@flesch.org}

\begin{abstract}
Announcing the release v2 of the MORX (Millions of Optical-Radio/X-ray Associations) catalogue which presents probable (40\%-100\% likelihood) radio/X-ray associations, including double radio lobes, to optical objects over the whole sky.  Detections from all the largest radio/X-ray surveys to June 2023 are evaluated, those surveys being VLASS, LoTSS, RACS, FIRST, NVSS, and SUMSS radio surveys, and \textit{Chandra}, \textit{XMM-Newton}, \textit{Swift}, and \textit{ROSAT} X-ray surveys.  The totals are 3\,115\,575 optical objects of all classifications (or unclassified) so associated.  The MORX v2 catalogue is available on multiple sites.
\end{abstract}

\keywords{catalogs --- x-rays: general --- radio continuum: general}


\section{Introduction} 
This is an update to the 2016 MORX catalogue of radio/X-ray associations onto optical objects \citep[MORXv1:][]{MORX}, using all the largest radio and X-ray source surveys available to 30 June 2023.  This update, version 2, has three times the number of optical objects so associated, and has a simpler format, just giving cumulative likelihoods and the radio/X-ray detection identifiers without details of their optical solutions, for ease of use.  Figure 1 shows MORXv2 coverage over the sky, and explains the features seen thereon.            

\begin{figure*} 
\includegraphics[scale=0.56, angle=0]{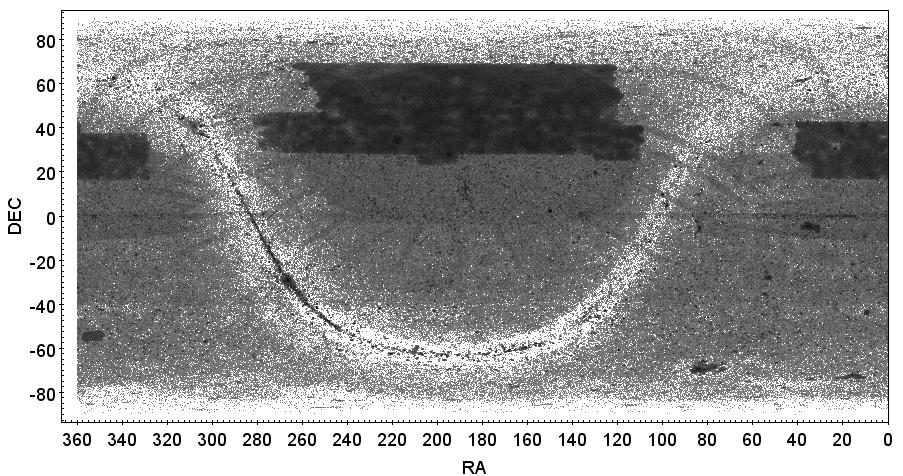} 
\caption{Sky coverage of MORX v2, darker is denser.  The Galaxy winds through with X-ray stars mostly, and the Magellanic Clouds are seen at lower right.  The dark footprints in the North show LoTSS radio coverage.  VLASS radio sources cover sky of $\delta>-40^{\circ}$ as does NVSS, RACS radio covers sky of $\delta<30^{\circ}$, and FIRST radio covers sky of $\delta>-10^{\circ}$ away from the Galaxy.  X-ray source coverage is broadly uniform over the sky, following where pointed investigations were done.  SDSS extensions radiate from their North sky optical coverage.  The thin strip along the equator shows the deep-sky investigations there.  The dense bead at lower left shows the XXL-South AAOmega X-ray field. \tiny{\textsl{(chart produced with TOPCAT \citep{TOPCAT})}}} 
\end{figure*}   

MORX v2 can be downloaded from CDS\footnote{\url{https://cdsarc.cds.unistra.fr/viz-bin/cat/V/158}} or from the MORX home page\footnote{\url{https://quasars.org/morx.htm}} which also provides a FITS file.  CDS and NASA HEASARC\footnote{\url{https://heasarc.gsfc.nasa.gov/W3Browse/all/morx.html}} provide query pages.  The MORX ReadMe gives essential information about the data, and the file "MORX-references.txt" gives the legend of citations for the data.  Many issues involved in the production of MORX are explained in the paper for MORXv1 \citep{MORX}, and the reader can consult there for in-depth topics.  Updates and changes are as presented below, starting with a listing of the input radio/X-ray catalogues for MORX v2.

\section{Input Radio/X-ray Catalogues}

In 2016 I published the Million Optical - Radio/X-ray Associations Catalogue \citep[MORXv1:][]{MORX} as a quick guide to all such associations able to be calculated from the largest optical, radio, and X-ray surveys over the whole sky.  That paper gives the methods used to produce this catalogue, including the optical field solutions and calculation of association likelihoods, and probabilistic classification of unclassified objects; the reader is referred there for those topics.  In the seven years since, there have of course been new editions of these surveys and new such surveys published; following is the list of input radio/X-ray surveys used for MORX v2.    

For radio sources there are the high-resolution VLASS, LoTSS, and FIRST surveys, the medium-resolution RACS survey, and the low-resolution NVSS and SUMSS surveys: 

\begin{itemize}
\item The Very Large Array Sky Survey \citep[VLASS:][]{VLASS} Quick Look catalogue\footnote{at \url{https://cirada.ca/catalogues}} is their first complete publication of VLASS radio sources.  VLASS covers the whole sky north of declination -40$^{\circ}$ in 3GHz to a depth\footnote{``depth'' = expected 67\% completeness at the given point source flux.} of 1mJy and astrometric accuracy of $\approx$0.5 arcsec in most places.  I accept only their Gaussian detections with S/N$\geq{4}$ and duplicate\_flag<2 and not flagged as probable sidelobes.  The VLASS Quick Look detection prefix of ``VLASS1QLCIR'' is here shortened to "VLA" for brevity.  

\item The LOFAR Two-metre Sky Survey second data release \citep[LoTSS:][]{LoTSS} which is a 120-168 MHz survey which covers 27\% of the Northern sky as is seen in dark on Figure 1.  It has a depth of 0.5 mJy and astrometric accuracy of 0.2 arcsec, and comprises 52\% of all core radio associations presented in MORX v2.  LoTSS comes in two primary tables, the Source `island' catalogue and the `Gaussian component' (i.e., functional detections) catalogue.  MORX processing is done on the detections catalogue which is architecturally many-to-one with the sources, although most sources have only the one detection.  The source catalogue provides source names prefixed with `ILT' (for International LOFAR Telescope), and that name is used for any MORX optically-associated detection which is the only detection for its source island.  If that detection is one of many for its source island, then LoTSS does not provide a name for it, so MORX constructs the detection name with the prefix `ILD' (for ILT Detection) and the J2000 of the detection location.  LoTSS is the largest input catalogue and took a week for me to process.     

\item The Faint Images of the Radio Sky at Twenty-cm survey catalogue \citep[FIRST:][]{FIRST} which is a 1.4GHz survey with a Northern-sky footprint away from the Galaxy, with a depth of 1mJy and astrometric accuracy of $\approx$1 arcsec. 

\item The Rapid ASKAP Continuum Survey \citep[RACS:][]{RACS} which is an 887.5 MHz survey which maps all sky south of declination +41$^{\circ}$, although this first catalogue of sources spans only ($-80^{\circ}<\delta<+30^{\circ}$) for quality reasons.  It has a point-source depth of 2 mJy and astrometric accuracy of $\approx$2 arcsec.  RACS uses the same source extraction software as LoTSS, so also presents a source catalogue and a `Gaussian' functional detections catalogue.  So, as with LoTSS, MORX uses the RACS-provided source name when an optically-associated detection is the only detection for its source, and otherwise constructs a detection name with the prefix `RACD' (for RACS Detection) and the J2000 of the detection location.  I did not use the small Galaxy RACS files which give only low-confidence associations due to optical crowding on the Galactic plane.        

\item The NRAO VLA Sky Survey catalogue \citep[NVSS:][]{NVSS} which covers the whole sky north of declination -40$^{\circ}$ in 1.4GHz to a depth of 2.5mJy and astrometric accuracy of $\approx$5 arcsec.

\item The Sydney University Molonglo Sky Survey catalogue \citep[SUMSS:][]{SUMSS} which is analagous to NVSS but covers the sky south of declination -30$^{\circ}$ in 843MHz.  It includes those associations marked as ``MGPS'' (Molonglo Galactic Plane survey) which is the Galactic plane component of the SUMSS survey.  
\end{itemize}

X-ray surveys are collected from pointed observations by X-ray detecting satellites, so their data are not characterized by completeness and depth as are the sky-sweeping radio surveys.  For X-ray sources there are 12 input catalogues from the high-resolution \textit{Chandra} and \textit{XMM-Newton} surveys, the medium-resolution \textit{Swift} survey, and lower-resolution XMM-Slew and \textit{ROSAT} satellite surveys; these are: 

\begin{itemize}
\item For \textit{Chandra} data, the \textit{Chandra} ACIS source catalog \citep[CXOG:][]{CXOG}, the \textit{Chandra} Source Catalog v1.1 \citep[CXO:][]{CXO} and v2.0 \citep[2CXO:][]{2CXO}, and the XAssist \textit{Chandra} source list \citep[CXOX:][]{CXOX}.  The CXO v1.1 data is preferred over the CXO v2.0 data because CXO v1.1 provides the observation ID for each detection, needed by MORX to calculate the optical solution, whereas CXO v2.0 provides stacked detections only.  \textit{Chandra} data has the best astrometric accuracy of the X-ray surveys, but published data is only to 2014.  CXO v2.1\footnote{\url{https://asc.harvard.edu/csc2/about2.1.html}} will publish more recent data when released, but may provide only stacked X-ray sources aligned to optical sources which, in my view, would prejudge the identification of true sources of X-ray emission.      

\item For \textit{XMM-Newton} data, catalogues used are the \textit{XMM-Newton} DR13 \citep[4XMM:][]{4XMM}, the \textit{XMM-Newton} DR3 \citep[2XMMi:][]{2XMM} which gives $\approx$20K valid clean detections dropped by its successors \citep[][Appendix D: ``mostly real sources"]{Rosen}, the XAssist \textit{XMM-Newton} source list \citep[XMMX:][]{XMMX}, and the \textit{XMM-Newton} Slew survey release 2 \citep[XMMSL2:][]{Slew}.  The Slew data is of much lower astrometric accuracy than the others, so is treated as a separate survey of low resolution in the MORX processing.  

\item \textit{Swift} data is taken from the Living \textit{Swift} X-ray Point Source catalogue \citep[SXPS:][]{SXPS} as at 01-July-2023.  The SXPS data are given both as observation-identified and stacked detections, so where both are present for a single source, MORX processing selects the observation-identified detections, as with the CXO processing.       

\item Lastly, the legacy \textit{ROSAT} data was of a lesser astrometric accuracy; catalogues used are the High Resolution Imager \citep[HRI:][]{HRI}, Position Sensitive Proportional Counter \citep[PSPC:][]{PSPC}, and the WGACAT \citep[WGA:][May 2000 edition]{WGA} which used different processing over the PSPC survey data.  The \textit{ROSAT} All-Sky Survey (RASS) is no longer used by MORX because its resolution was too coarse to yield confident optical associations in isolation.  
\end{itemize}

Counts of associations from all these input catalogues are given in Table 1, plus a count of ``primary associations'' which are one-to-one with optical sources for each of radio and X-ray.  The radio associations are calculated independently from the X-ray associations; an optical source showing both radio and X-ray associations has earned each one separately, and so can tally both as a primary radio association and a primary X-ray association in Table 1.

\begin{table} 
\caption{Summary of Radio/X-ray Associations in MORX}
\begin{tabular}{@{\hspace{0pt}}l@{\hspace{0pt}}rrr}
\hline 
Source       & \# total core & \# primary$^{*}$ core & \# double \\
Catalogue(s) & associations  & associations         & radio lobes \\
\hline 
Radio surveys \\
VLASS  &  439283 &  439283 & 15763  \\
LoTSS  & 1804886 & 1710297 & 73142  \\
FIRST  &  275552 &   69766 &  9000  \\
RACS   &  582668 &  337540 & 12009  \\
NVSS   &  316039 &   78759 &   675  \\
SUMSS  &   47549 &   15891 &    42  \\
total Radio Surveys & 3465977 & 2651536 & 110631 \\
\hline 
X-ray Surveys \\ 
CXOG          &  63503 &  63503 &  \\
CXO v1.1      &  13376 &  13376 &  \\
CXO v2.0      &  6807  &   6807 &  \\
CXOX          &  18966 &  18966 &  \\
\textit{total Chandra}  & \textit{102652} & \textit{102652}  &  \\
4XMM-DR13     & 250592 & 221701 &  \\
2XMM-DR3      &   4541 &   4086 &  \\
XMMX          &   8516 &   8040 &  \\
\textit{total XMM-Newton} & \textit{263649} & \textit{233827} & \\
\textit{Swift} LSXPS & 120647 & 95484 &  \\
XMM Slew v2.0        &  11428 &  8778 &  \\
\textit{ROSAT} HRI   &  11177 &  5146 &  \\
\textit{ROSAT} PSPC  &  17813 & 12155 &  \\
\textit{ROSAT} WGA   &   2760 &  2321 &  \\
\textit{total ROSAT}  &  \textit{31750} & \textit{19622} &  \\
total X-ray Surveys & 530126 & 460363 & \\
\hline 
\multicolumn{4}{@{\hspace{0pt}}l}{$^{*}$ For each radio source, the `primary' association is the first} \\
\multicolumn{4}{@{\hspace{0pt}}l}{occurence on this list from top to bottom, e.g., if a radio source} \\
\multicolumn{4}{@{\hspace{0pt}}l}{has associations from the RACS and NVSS surveys, then RACS} \\
\multicolumn{4}{@{\hspace{0pt}}l}{has the primary association because it is topmost on this list.} \\
\multicolumn{4}{@{\hspace{0pt}}l}{X-ray sources are tallied similarly.} \\
\end{tabular}
\end{table}

\section{The Optical Background used in MORX v2}

The optical astrometry for MORX v2 hails from the ``best" sources which comprise a mix of 23.7\% \textit{Gaia}-EDR3 \citep{Gaia3}, 46.2\% Pan-STARRS \citep{PS}, 19.0\% SDSS-Sweeps\footnote{at \url{https://data.sdss.org/sas/dr9/boss/sweeps/dr9/}} supplemented with SDSS XDQSO \citep{XDQSO} optical data, 0.6\% from DES \citep{DES} or discovery authors' papers, and 10.5\% from the 0.1-arcsec resolution All-Sky Portable optical catalogue \citep[ASP:][]{ASP} which includes legacy optical data.  Most MORX objects are optically too faint for \textit{Gaia} coverage, but SDSS is used as available, Pan-STARRS gives almost complete coverage for $\delta>-30^{\circ}$, and ASP covers south of there.    

By contrast, for optical photometry MORX gives only 2 bands, red and blue, so no attempt is made to give "best" values for these, as researchers will source those elsewhere, anyway.  Instead, MORX presents historically-valuable calibrated POSS-I (1950's epoch) \& POSS-II/UKST (1980's epoch) photometry sourced from ASP.  In particular, the priority is to present POSS-I magnitudes because the blue POSS-I \textsl{O}, centred at violet 4050\AA, is well-separated from the red Cousins 6400\AA, and the two plates were always taken on the same night thus giving accurate red-blue colour even for variable objects.  In total, MORX gives photometry comprised of 37.4\% POSS-I, 26.7\% POSS-II/UKST, 31.6\% SDSS-Sweeps, 1.9\% Pan-STARRS, 0.6\% DES, and 1.9\% other.   

For MORX v2 (this edition), I have modified the optical field solutions documented in MORXv1, in that I have constrained the optical field shifts to be no more than 8 arcsec in each of RA and DEC.  This is because of the concise presentation of this edition where each radio/X-ray association is displayed simply with its J2000-based name, without elaboration.  Thus each such association needs to be ``seen to be" correct, lest it be ``seen to be" wrong.  Thus I drop any association found to be positionally too far removed from its own J2000-based name (that name showing its original astrometry, in principle); in practice, this drops about 2\% of the X-ray associations but does not disturb the radio associations.  Spot checks on the removed X-ray field shifts show most were only marginally aligned to the optical background, but some were quite clear, e.g., the \textit{ROSAT}-HRI field US700009H.N1 (centred over NGC 3628) shifted 15 arcsec which revealed X-ray emission for 5 quasars in a closely-aligned fit.  However, usually such places have since been resurveyed anyway, thus clarifying the true optical sources, as in this case which was resurveyed by \textit{Chandra} which confirmed the field-shifted HRI-found associations, and so preserving those optical sources in MORX.

\section{Association Likelihoods of Radio/X-ray Sources}

The calculations of the likelihood of core radio or X-ray associations to an optical object are as presented in MORX v1, but now those likelihoods are calculated in 0.1-arcsec offset bins instead of the previous 1-arcsec bins.  Smoothing is now done onto a logarithmic profile instead of linear, to avoid over-representing likelihoods.  While the likelihood lower-limit cutoff for MORX remains at 40\%, past analysis has shown performance at that lower limit to be better than that; a Y2009 test of this against the predecessor QORG catalogue \citep{QORG} is given at \url{https://quasars.org/docs/Testing-QORG-via-SDSS-DR7.txt} .    

The best quasar candidates (66\,026 with pQSO$\approx{99}$\% in bulk) are also presented in the Milliquas v8 catalogue \citep{MQ} which was extracted alongside MORX out of a large frozen database holding data to 30 June 2023.  Milliquas used to hold many more quasar candidates; those are now included here in this MORX v2 catalogue.

\section{Comparison Against L\symbol{111}TSS Optical Identifications}

A few weeks after this MORX paper was first submitted, a large optical identifications catalogue for LoTSS-DR2 radio sources appeared \citep[][hereinafter: LOPTS]{Lopts} which contains 3\,519\,018 optical identifications (952 more than reported in their Table 5) over 4\,116\,934 LoTSS sources for a hit rate of 85.48\%.  MORX, on the other hand, shows ``only" 1\,829\,471 optical identifications for LoTSS sources.  It behoves us to cross-check these catalogues to see what can be learned about the nature of their data.

Three substantive differences between the two catalogues' handling of the LoTSS radio data are:

\begin{itemize}
\item LOPTS uses optical data wholly from the DESI-DR9 \citep{DESIopt} ``sweeps" catalogues whilst MORX uses the optical data outlined in Section 3.  DESI photometry reaches 1-2 magnitudes deeper than surveys used by MORX, so LOPTS will have more optical objects to match to.  The DESI data covers most but not all of the LoTSS-covered sky (which is $\approx$14\% of all-sky), so the remaining $\approx$300 deg$^{2}$ of LoTSS sky is not covered by LOPTS, but MORX does cover it, being an all-sky catalogue.   

\item LOPTS matches opticals primarily to the LoTSS `source' table (described in Section 2, above), whilst MORX matches only to the LoTSS `Gaussian' (detections) table.  However, often a radio source manifests as multiple detections, 2 or 3 in the case of double lobes, more for extended or saturating emission.  Thus LOPTS supplements the groupings provided by the source extraction software with further visually-ascertained groupings.  In contrast, MORX groups only raw detections via algorithms previously used by MORXv1 and earlier works, including algorithmic identification of double lobes.  All this will impact optical selection.   

\item Both catalogues give a confidence-of-association figure, MORX as a simple percentage, LOPTS as a Likelihood Ratio (LR) which converts to confidence-of-association by conf = LR/(1+LR).  MORX gives optical identifications down to a confidence of 40\%, and no lower.  LOPTS reports down to about 24\% confidence (i.e., LR=0.3) for those not identified visually.  (MORX has no flagged visual identifications, but thousands of visual checks have been done over the life of the catalogue development and use.)  
\end{itemize}
 
Let's be clear that not all these identifications are equal; if the reported confidence-of-associations are accurate, then one 100\%-confidence association is worth two 50\%-confidence associations.  If the confidence pcts are totalled, then MORX will deliver 1\,695\,614 true associations out of 1\,829\,471 reported associations, and LOPTS will deliver 3\,246\,715 true associations out of 3\,451\,555 reported associations for which an LR is given.  This confidence-driven ``yield" is an essential part of any probabilistic analysis, and is presented alongside count numbers below and in Table 2.     

Now let's see how well the catalogues agree, the better to gauge any points of disagreement.  Matching by LoTSS source name and accepting only those where the MORX and LOPTS optical co-ordinates are within one arcsec of each other (although arbitrarily far from the radio source), we obtain 1\,494\,158 agreeing radio-optical associations.  For these, MORX gives a mean confidence-of-association of 95.084\%, while LOPTS gives 97.268\%.  Thus MORX is more pessimistic, but the data-driven MORX calculations evaluate and so incorporate the contribution of unseen background sources.  Since LOPTS uses deeper DESI optical data, the chance of unseen background sources is less, so LOPTS is justified to display more confidence.  Also, over these data, the mean of the absolute difference between the MORX and LOPTS confidence-of-association is 4.244\%, and the median absolute difference is just 1.332\%.  Thus the two catalogues align pleasingly well, for those LoTSS-optical associations that they share.

\begin{table*} 
\scriptsize 
\caption{Optical selections for LoTSS sources -- MORX vs LOPTS (refer Section 5)}
\begin{tabular}{lrr@{\hspace{24pt}}rrrr}
\hline 
\ & \\
\multicolumn{7}{c}{MORX Counts \& Confidence-Based Yields} \\
\ & \\
 & \multicolumn{2}{l}{IN DESI FOOTPRINT} & \multicolumn{2}{c}{OUTSIDE OF DESI} & \multicolumn{2}{c}{TOTAL} \\
 & Counts & Yields & Counts & Yields & Counts & Yields \\
\hline
\multicolumn{7}{l}{MATCHES ACROSS MORX AND LOPTS (opticals within 2 arcsec of eachother):} \\
Isolated Sources from both MORX \& LOPTS & 1441336 & 1372317 & 6801 & 6393 & 1448137 & 1378710 \\
MORX isolated sources / LOPTS blobs-lobes & 58577 & 51359 &  358 &  307 & 58935 & 51666 \\
MORX cored lobes$^{\dagger}$ / LOPTS isolated sources & 25892 & 24912 & 89 & 84 & 25981 & 24996 \\ 
MORX blank lobes$^{\dagger}$ / LOPTS isolated sources & 1541 & 1295 & 10 & 8 & 1551 & 1303 \\
MORX cored lobes / LOPTS blobs-lobes & 9046 & 8311 & 64 & 58 & 9110 & 8369 \\
MORX blank lobes / LOPTS blobs-lobes & 3734 & 3140 & 25 & 20 & 3759 & 3160 \\ 
\textit{TOTAL MORX-LOPTS MATCHES:} & \textit{1540126} & \textit{1461334} & \textit{7347} & \textit{6870} & \textit{1547473} & \textit{1468204} \\
\hline 
\multicolumn{7}{l}{MORX OPTICALS VALID AND UNMATCHED TO LOPTS:} \\
Isolated Sources & 85466 & 70039 & 59711 & 54033 & 145177 & 124072 \\
Cored Lobes & 10968 & 9869 & 1854 & 1671 & 12822 & 11540 \\
Blank Lobes & 9037 & 6750 & 804 & 591 & 9841 & 7341 \\
\textit{TOTAL MORX VALID UNMATCHED:} & \textit{105471} & \textit{86658} & \textit{62369} & \textit{56295} & \textit{167840} & \textit{142953} \\
\hline 
\multicolumn{7}{l}{OPTICALS DISAGREE, LOPTS OPTICAL SELECTED OVER MORX$^{*}$ (opticals 2+ arcsec apart):} \\ 
Isolated Sources & 103359 & 76333 & 721 & 560 & 104080 & 76893 \\
Cored Lobes & 3252 & 2726 & 36 & 29 & 3288 & 2755 \\
Blank Lobes & 6770 & 4794 & 20 & 15 & 6790 & 4809 \\
\textit{TOTAL MORX SUPERSEDED:} & \textit{113381} & \textit{83853} & \textit{777} & \textit{604} & \textit{114158} & \textit{84457} \\
\hline
TOTAL MORX OPTICAL SELECTIONS FOR LoTSS: & 1758978 & 1631845 & 70493 & 63769 & 1829471 & 1695614 \\
\hline
\hline
\ & \\
\multicolumn{7}{c}{LOPTS Counts \& Confidence-Based Yields (no-LR visuals counted as 100\% confidence)} \\
\ & \\
 & \multicolumn{2}{l}{IN DESI FOOTPRINT} & \multicolumn{2}{c}{OUTSIDE OF DESI} & \multicolumn{2}{c}{TOTAL} \\
 & Counts & Yields & Counts & Yields & Counts & Yields \\
\hline
\multicolumn{7}{l}{MATCHES ACROSS LOPTS AND MORX (opticals within 2 arcsec of eachother):} \\
Isolated Sources from both LOPTS \& MORX & 1441336 & 1404857 & 6801 & 6579 & 1448137 & 1411436 \\
LOPTS isolated sources / MORX lobes  & 27433 & 26742 & 99 & 95 & 27532 & 26837 \\
LOPTS blobs-lobes / MORX isolated sources & 58577 & 54095 & 358 & 306 & 58935 & 54401 \\ 
LOPTS blobs-lobes / MORX lobes & 12780 & 12653 & 89 & 83 & 12869 & 12736 \\
\textit{TOTAL LOPTS-MORX MATCHES:} & \textit{1540126} & \textit{1498347} & \textit{7347} & \textit{7063} & \textit{1547473} & \textit{1505410} \\
\hline 
\multicolumn{7}{l}{LOPTS OPTICALS VALID AND UNMATCHED TO MORX:} \\
Isolated Sources with Likelihood Ratio (LR): & 1758919 & 1624109 & 10084 & 9292 & 1769003 & 1633401 \\
Visual Isolated Sources without LR: & 37140 & 37140 & 22 & 22 & 37162 & 37162 \\
Blobs / Lobes with LR: & 123793 & 100376 & 1211 & 919 & 125004 & 101295 \\
Visual Blobs / Lobes without LR: & 5551 & 5551 & 5 & 5 & 5556 & 5556 \\
\textit{TOTAL LOPTS VALID UNMATCHED:} & \textit{1925403} & \textit{1767176} & \textit{11322} & \textit{10238} & \textit{1936725} & \textit{1777414} \\
\hline 
\multicolumn{7}{l}{DISAGREEMENT, LOPTS OPTICALS SUPERSEDED BY MORX DOUBLE LOBES (in this exercise):} \\ 
Isolated Sources & 32054 & 29564 & 91 & 83 & 32145 & 29647 \\
Blobs / Lobes & 2649 & 1566 & 26 & 13 & 2675 & 1579 \\
\textit{TOTAL LOPTS SUPERSEDED:} & \textit{34703} & \textit{31130} & \textit{117} & \textit{96} & \textit{34820} & \textit{31226} \\
\hline
TOTAL LOPTS OPTICAL SELECTIONS FOR LoTSS: & 3500232 & 3296653 & 18786 & 17397 & 3519018 & 3314050 \\
\hline
\hline 
\ & \\
\multicolumn{7}{l}{$^{*}$ this decision-branch used for this exercise.  There are a mix of objects on each side of the 2-arcsec dividing line,} \\ 
\multicolumn{7}{@{\hspace{12pt}}l}{but visual inspection of all those objects is beyond the scope of this paper.} \\
\multicolumn{7}{l}{$^{\dagger}$ ``cored lobes": double radio lobes with a radio centroid. ``blank lobes": same but no radio centroid in the data.} \\
\ & \\
\ & \\
\end{tabular}
\end{table*}

\begin{table*} 
\caption{Sample lines from the MORX catalogue (left half placed on top of right half)} 
\includegraphics[scale=0.3, angle=0]{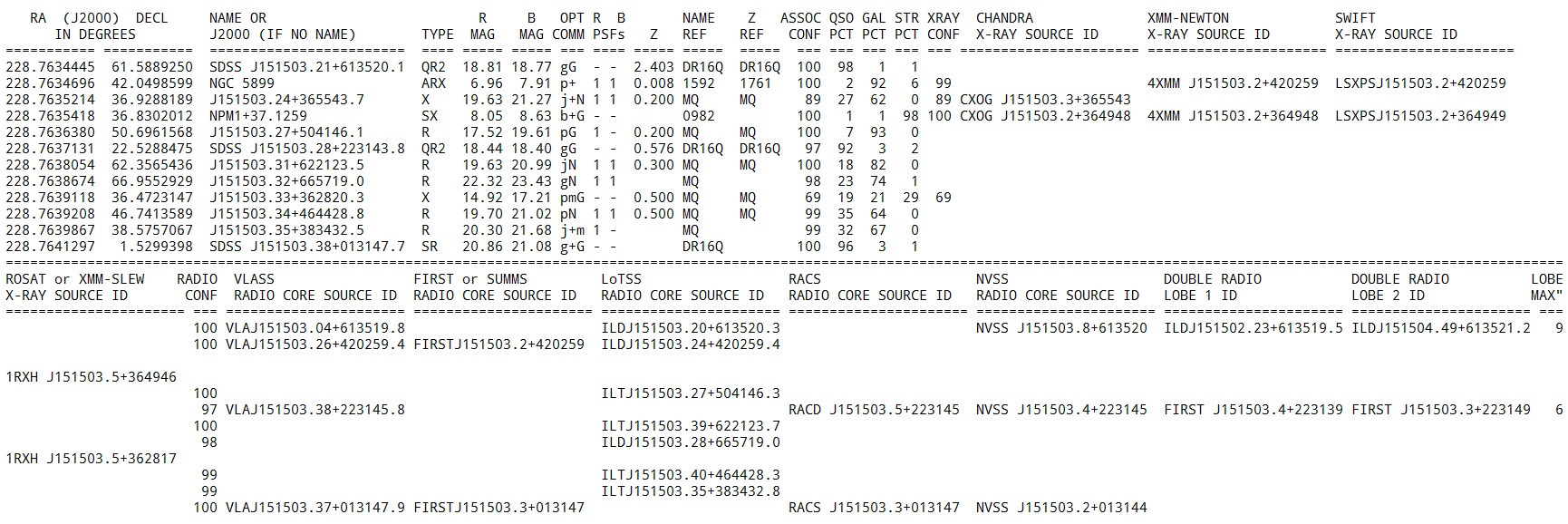} 
\tiny 
 \\
 \\
Notes on columns (see ReadMe for full descriptions): 
\begin{itemize}
  \item TYPE:  R=core radio detection, X=X-ray detection, 2=double radio lobes (calculated), Q=QSO, A=AGN, G=galaxy, S=star. 
  \item REF \& ZREF: citations for name and redshift; citations are indexed in the file "MORX-references.txt".
  \item OPT COMM:  comment on photometry: p=POSS-I magnitudes so blue is POSS-I \textsl{O}, j=POSS-II/UKST \textsl{Bj}, g=SDSS \textsl{g} \& \textsl{r}, +=optically variable, m=nominal proper motion, G=\textit{Gaia} astrometry, N=Pan-STARRS astrometry.
  \item R/B PSFs:  '-'=stellar, 1=fuzzy, n=no PSF available, x=not seen in this band.
  \item ASSOC/RADIO/X-RAY CONF:  calculated percentage confidence that this source is truly associated to this optical object in total/radio/X-ray, respectively. 
  \item QSO/GAL/STR PCT:  based on its photometry and the radio/X-ray association(s), the calculated percentage confidence that this optical object is a QSO/galaxy/star. 
  \item LOBE MAX":  offset of the longest radio lobe from the centroid, in arcsec.  This calculation uses detection centroids, so visual lobe length can be longer. 
\end{itemize} 
The full table can be downloaded from \url{http://quasars.org/morx.htm}, also available in FITS. \\
 \\
 \\
 \\
\end{table*}

With the two catalogues thus showing equivalent throughputs, we now list where they differ.  Counts are supplemented with the yields after factoring in confidence-of-association: 
\begin{itemize} 
\item LOPTS has 1\,936\,725 LoTSS-optical associations not presented by MORX.  Of these, 42\,718 are visually evaluated without an LR, so crediting those as 100\% confident, the total yield of confidences is 1\,777\,414 LOPTS-only associations.  As we have established the similar performance of LOPTS and MORX above, the expectation is that the LOPTS LRs are assessing these with good accuracy.  Inspection of those data show dominantly faint opticals, assuredly from DESI, which LOPTS carries and MORX does not.      
    
\item MORX has 167\,840 (yield 142\,953) valid LoTSS-optical associations not in LOPTS.  Of those, 62\,369 (yield 56\,295) are outside of the DESI footprint, MORX being an all-sky catalogue.  I visually inspected some within the DESI footprint to find why LOPTS did not match those.  Among a variety of situations, some stood out: (1) Galaxies can have multiple LoTSS sources not consolidated by the source extraction software, so sometimes LOPTS and MORX select different nominal radio sources for the same optical association, thus evading the matching method of this exercise, (2) Some faint reddish galaxies appear to be in Pan-STARRS or SDSS data but not in the DESI data, and (3) When a LoTSS source has multiple components (detections), one of those components may match well to a prominent optical, but somehow this can get missed by the LOPTS method.  An example is the LoTSS source ILTJ000412.62+325711.8 which has a component MORX-named as ILDJ000412.17+325709.5, which matches with 100\% confidence to the SDSS-DR16Q quasar SDSS J000412.15+325709.2, z=1.386 -- but missed by LOPTS.  Browsing the LOPTS components table shows this component to be absent, with only the other one component reported for this source, for some reason.  Querying the whole LOPTS components table finds 28\,841 sources with only one component reported, curiously.  A more thorough evaluation of this overall topic is out of scope for this paper.     

\item MORX uses a unique algorithm to identify double radio lobes, including ones without radio centroids which require visual follow-up.  LOPTS finds these from the source extraction software, and visual evaluation.  In general, double radio lobes need a radio centroid confidently associated to an optical object, to be regarded as fully confirmed (hereinafter: ``cored lobes").  The MORX-presented lobes without a radio centroid (hereinafter: ``blank lobes") are assigned onto the ``best" candidate optical centroid; visual inspection there often reveals a faint radio signature which confirms those lobes, but sometimes the faint radio signature (if one is found at all) is onto another nearby optical, thus falsifying the MORX candidate.  In this exercise, as shown in Table 2 (4$^{th}$ and 6$^{th}$ lines of data), 5310 blank MORX lobes match their optical-only centroids to LOPTS radio-optical sources which thus provide radio-optical centroids for those MORX lobes, provisionally converting them from ``blank lobes" to ``cored lobes".  
    
Of 32\,741 MORX-only LoTSS double lobes, 10\,078 get falsified in this exercise by a lobe radio source matching to a LOPTS optical association, because the faint DESI optical was not in the MORX optical data.  But also 34\,820 LOPTS associations get superseded (in this exercise) by qualifying MORX double lobes, either because the lobes have a supporting core radio-optical association which yields a high-confidence lobe configuration, or because the LOPTS optical was offset $\geq$2 arcsec from the radio detection, thus confirming that radio detection as a lobe signature, in this exercise.  
\end{itemize}

Thus, this exercise has shown the MORX and LOPTS catalogues to be well-aligned in their throughput, with LOPTS having twice as many LoTSS-optical associations due to its use of deep DESI optical data absent from MORX, whilst MORX associations are generally confirmed by LOTSS where not overtaken by those same deep DESI opticals.  There are of course some disagreeing optical selections inevitable in such large data.  The one contentious arena is that of double radio lobes, which can be settled only by individual visual inspections.
        
\section{Catalogue Layout}

The MORX catalogue presents one line per optical object; Table 3 displays 12 sample lines which are wrapped with the left half of the block of lines shown on top of the right half.  Names are given as found in the literature; else, if anonymous, the J2000 location is given as a convenience to the user.  The ReadMe gives full details and indexes to the columns.

\section{Conclusion} 

The MORX v2 catalogue is presented which gives 3\,115\,575 probable radio/X-ray associations onto optical objects over the whole sky, including double radio lobes, using all the largest radio/X-ray source surveys to June 2023.  Identifications are included to provide an informative map for pointed investigations.

\section*{Acknowledgements}
Thanks to Heinz Andernach for steering me onto VLASS, RACS, and LoTSS, although processing the latter almost broke my computer.  Also thanks to the referee for good suggestions.  This work was not funded.

\section*{Data Availability}

MORX v2 can be downloaded from CDS at \url{https://cdsarc.cds.unistra.fr/viz-bin/cat/V/158} or from its home page at \url{https://quasars.org/morx.htm} which also provides a FITS file.  Both sites also provide the ReadMe and the references list.  Query pages are provided by CDS and NASA HEASARC at \url{https://heasarc.gsfc.nasa.gov/W3Browse/all/morx.html}.


\label{lastpage}
\end{document}